# Moiré Band Topology in Twisted Bilayer Graphene

Chao Ma,[†] Qiyue Wang,[‡] Scott Mills,[§] Xiaolong Chen,[†#] Bingchen Deng,[†] Shaofan Yuan,[†] Cheng Li,[†] Kenji Watanabe,[∥] Takashi Taniguchi,[∥] Du Xu,[*§] Fan Zhang,[*‡] and Fengnian Xia[*†]

[†]Department of Electrical Engineering, Yale University, New Haven, Connecticut 06511, USA

[‡]Department of Physics, The University of Texas at Dallas, Richardson, TX 7508, USA

[§]Department of Physics and Astronomy, Stony Brook University, Stony Brook, NY11794, USA

[∥]National Institute for Materials Science, 1-1 Namiki, Tsukuba 305-0044, Japan

## ABSTRACT

Recently twisted bilayer graphene (t-BLG) emerges as a new strongly correlated physical platform near a magic twist angle, which hosts many exciting phenomena such as the Mott-like insulating phases, unconventional superconducting behavior and emergent ferromagnetism. Besides the apparent significance of band flatness, band topology may be another critical element in determining strongly correlated twistronics yet receives much less attention. Here we report compelling evidence for nontrivial noninteracting band topology of t-BLG moiré Dirac bands through a systematic nonlocal transport study, in conjunction with an examination rooted in K-theory. The moiré band topology of t-BLG manifests itself as two pronounced nonlocal responses in the electron and hole superlattice gaps. We further show that the nonlocal responses are robust to the interlayer electric field, twist angle, and edge termination, exhibiting a universal scaling law. While an unusual symmetry of t-BLG trivializes Berry curvature, we elucidate that two $Z_2$ invariants characterize the topology of the moiré Dirac bands, validating the topological edge origin of the observed nonlocal responses. Our findings not only provide a new perspective for understanding the emerging strongly correlated phenomena in twisted van der Waals heterostructures, but also suggest a potential strategy to achieve topologically nontrivial metamaterials from topologically trivial quantum materials based on twist engineering.

It is widely known that overlaying two identical periodic lattices with a relative twist generates a larger-scale interference structure, i.e., the moiré pattern. For two-dimensional (2D) materials, such twists create moiré superlattices by reducing translation symmetry in real space and folds electron Bloch bands into moiré Brillouin zones (MBZ) in momentum space.[1-5] The moiré bands can exhibit striking phenomena such as the Hofstadter's butterfly in fractal quantum Hall effect[6-8] and the moiré potential-modulated interlayer excitons.[9-12] Remarkably, strongly correlated electron behavior including Mott-like insulating phases and possibly unconventional superconductivity have been discovered in twisted bilayer graphene[13-17] (t-BLG) near a magic angle[4] (~1.1°) and in aligned trilayer graphene/hexagonal boron nitride (hBN) heterostructures.[18,19] Besides their extreme band flatness, nontrivial band topology may be another crucial feature in determining the observed strongly correlated phases. One celebrated example is the well-studied incompressible fractional quantum Hall effect,[20] in which the partially filled Landau level has a nontrivial Chern number.[21] Another example is the predicted $Z_4$ parafermions,[22] for which the $8\pi$ Josephson effect is mediated by a helical edge state of 2D topological insulator (TI)[23] with a sufficiently small velocity. Therefore, it is of significance to investigate the possible topology of the active moiré bands in these correlated systems. Although topological features such as nontrivial Wilson loop spectral flows have been obtained theoretically[24-26] in t-BLG, there is no experimental evidence yet to support any nontrivial band topology in t-BLG.

Here we experimentally reveal the universal and unique moiré band topology of t-BLG at small twist angles by using systematic nonlocal transport measurements.[27-31] Previously, a nonlocal measurement scheme has been employed in detecting the helical edge state of 2D TI[27] and the topological valley current driven by Berry curvature.[28-31] In this work, pronounced nonlocal responses are observed in both the electron and hole superlattice-induced bandgaps (four electrons or holes per unit cell) of t-BLG[32] and consistently observed in more than 15 t-BLG devices with twist angles between ~1.3° and ~1.9°, showing their robustness to twist angle and edge termination. Moreover, the nonlocal responses persist in a wide range of applied displacement fields from 0 to 0.62 V/nm. However, the nonlocal responses disappear in t-BLG with twist angles of ~0.75° and ~0.42°, in which the superlattice bandgaps vanish. We elucidate that the unique $C_{2z}T$ symmetry of t-BLG, though trivializing any Berry curvature, gives rise to two nontrivial bulk $Z_2$ invariants that previously appear in six and seven dimensions in the celebrated periodic table of topological classification.[33] While one invariant protects the moiré Dirac points, the other dictates the presence of one pair of counter-propagating edge states per spin-valley in each superlattice gap. We further show that the observed non quantization and universal scaling of nonlocal resistance are consistent with the $C_{2z}T$ and valley symmetry breaking at edges. Our findings not only unveil the appealing moiré band topology of t-BLG but also may offer a universal pathway for creating topological metamaterials by twist engineering 2D materials. The discovered moiré band topology may provide a new perspective for deciphering the tantalizing strongly correlated phenomena[13-17, 34] in magic-angle t-BLG.

**RESULTS AND DISCUSSION**

A schematic of our typical t-BLG devices is shown in top panel of Figure 1a. We employed a "tear-and-stack" technique to fabricate hBN encapsulated t-BLG heterostructures as reported previously(Methods).[5, 32, 35] A Hall-bar geometry was defined through reactive ion etching, and electrical contacts to t-BLG were made at the edge.[35] The total carrier concentration $n$ in t-BLG is controlled by tuning the voltage applied to the global silicon back gate. The schemes of local and nonlocal measurements are illustrated in the lower panels in Figure 1a. We first characterize the t-BLG moiré superlattice by measuring the four-probe local resistance $R_L$ as a function of $n$ in two devices (D1 and D2) shown in the insets in Figures 1b and 1c.

Three resistance maxima are observed at the charge neutrality point and at $n = \pm n_s \approx \pm 6.50 \times 10^{12} \text{cm}^{-2}$ (Figure 1b) for D1 at 80 K, at which the Fermi level crosses the degenerate Dirac points from the two original graphene monolayers and the two moiré superlattice-induced band gaps due to the interlayer hybridizations,[32] respectively. Here, $n_s/n_0 = 4$, where $n_0$ corresponds to one electron per moiré unit cell,[32] from which we can deduce the twist angle of this device to be ~1.68°. The error is estimated to be around ±0.05°. The superlattice gaps are extracted to be ~40 meV based on the temperature dependent transport measurements (See Figure S1a and S1b for the complete temperature dependence of $R_L$ and superlattice bandgap extraction). Strikingly, pronounced nonlocal responses $R_{NL} = V_{NL}/I$ are observed in the superlattice gaps on both the electron- and hole-sides, as featured in Figure 1b at 80 K. The peaks of $R_{NL}$ have a narrower range in $n$ compared with that of $R_L$ and attenuate to zero away from $\pm n_s$. The $R_{NL}$ at lower temperature is discussed and presented in Supporting Information Section 1 and Figure S1c.

We first exclude the possibility of the observed nonlocal resistance in D1 being an Ohmic contribution induced by stray-currents, where the Ohmic nonlocal resistance can be estimated by the van der Pauw relation[28-30, 36-39] $R_{NL,Ohmic} = R_L w/\pi L \exp(-\pi L/w)$. Here $L = 3$ μm is the channel length between the driving and probing terminals, and $w = 1.2$ μm is the channel width in D1. The Ohmic contribution only yields $R_{NL,Ohmic} < 1\Omega$ at the superlattice gaps, which is at least two orders of magnitude smaller than our observed $R_{NL}$ peak values. In addition, experimental artifacts may arise from the diffusive charge transport in an imperfect device, e.g., a Hall bar device with a pair of unintentionally misaligned side terminals. Although such experimental artifacts can vary from device to device, they should produce a nonlocal resistance proportional to the local resistance in the same device. However, pronounced nonlocal resistances are only observed in the two superlattice gaps but not at the Dirac point (charge neutrality), as shown in Figure 1b, although the local resistances at the superlattice gaps and the Dirac point only differ by a few times. Were the nonlocal signals at the superlattice gaps due to the experimental artifacts, we should have observed a similarly strong nonlocal resistance peak at the Dirac point. In fact, as discussed below, we do not observe meaningful nonlocal resistance at the Dirac point for all our t-BLG devices with twist angles between ~1.3° to ~1.9°. Moreover, experimental errors due to the measurement set-ups are estimated to be negligible compared with the experimentally observed nonlocal signals at the superlattice gaps (Supporting Information Section 2 and Figure S2). Finally, we also exclude the trivial mechanisms such as the band-bending-induced metallic edge conduction based on the thermal activation behavior and the exponential length dependence of the nonlocal responses that we shall present below. Therefore, the observed strong nonlocal signals are intrinsic properties that are related to the superlattice gaps of t-BLG at $n_s/n_0 = 4$.

As for the device with a twist angle of ~0.75° (D2 measured at 80 K, Figure 1c), due to the reduced energy dispersion of the moiré band,[13, 40, 41] resistance maxima at the charge neutrality and other positions are thermally smeared at 80 K. By lowering the temperature down to 2 K, we observe two insulating peaks at $n = \pm 2.60 \times 10^{12} \text{cm}^{-2}$ and an additional peak at $n = +1.31 \times 10^{12} \text{cm}^{-2}$ that has a metallic behavior (Figure S3a). In the Landau level fan diagram (Figure S3b), Landau level fans emanating from both sides of the charge neutrality point and of $n = 1.31 \times 10^{12} \text{cm}^{-2}$ can be clearly resolved. Based on the Wannier theory,[13-17, 34] we assign $1.31 \times 10^{12} \text{ cm}^{-2}$ to $n/n_0 = +4$ and estimate the twist angle to be 0.75°. As such, our observation of the insulating states at $n/n_0 = \pm 8$ is also consistent with previous works,[13, 40] where these unusual insulating behavior was attributed to strong electron correlations. In this device, the strong nonlocal resistance is absent at all carrier concentrations at 80 K (Figure 1c); in fact, the nonlocal resistance is always on the order of the Ohmic contribution down to 2 K (Supporting Information Section 3 and Figure S3c). Moreover, for ~0.42° t-BLG (device D3), local resistance peaks appear at single-particle band fillings without any thermal activation behaviors and no significant nonlocal responses are observed at all temperatures (Supporting Information Section 4 and Figure S4c ).

In order to identify the physical origin of the strong nonlocal signals in D1, we performed temperature dependent measurements. With lowering the temperature to 1.7 K, $R_{NL}$ continues to grow and exhibits fluctuations without any obvious quantization in D1 (Figure S1c), which excludes the presence of protected gapless edge states like those in topological insulators.[27] In addition, theoretical analysis (Supporting Information Section 5) shows that Berry curvature vanishes in t-BLG, which precludes the possibility of valence-band Berry curvatures[42]-driven nonlocal transport that was studied in several 2D materials.[28-31]

We now turn to the intrinsic band properties of t-BLG. Figures 2a and 2b show the calculated moiré band structures of 1.68° and 0.75° t-BLG (Methods), respectively. The difference between these two cases is evident. In Figure 2a, the red region in the middle and the two white regions adjacent to it highlight the two moiré Dirac bands (per spin-valley) and the two superlattice gaps, respectively. On the contrary, the superlattice gaps close in Figure 2b, consistent with a previous report.[40] It is clear that the existence of superlattice gaps is critical for the observation of the strong nonlocal signals. For t-BLG with superlattice gaps at $n/n_0 = \pm 4$, the presence of a nontrivial topological invariant for the entire moiré Dirac bands is possible. For t-BLG without the superlattice gaps, the moiré Dirac bands are no longer energetically isolated, and such a topological property becomes irrelevant. Below we use symmetry and *K*-theory to elucidate the nontrivial band topology of t-BLG with superlattice gaps at $n/n_0 = \pm 4$ and show that this unprecedented property gives rise to delocalized topological edge modes responsible for the observed strong nonlocal signals.

In t-BLG, because $(C_{2z}T)H(\mathbf{k})(C_{2z}T)^{-1} = H(\mathbf{k})$ and $(C_{2z}T)^2 = 1$, the Hamiltonian $H(\mathbf{k})$ belongs to class AI in the celebrated periodic table of topological classification rooted in *K*-theory.[33] In this classification (Table 1), the topological invariant of $H(\mathbf{k})$ is determined by the effective dimensions *d-D*, where *d* (*D*) denotes the number of dimensions in which momentum is odd (even) under $C_{2z}T$. Evidently, t-BLG can be characterized by two independent $Z_2$ invariants, as $H(\mathbf{k})$ is featured by *d*=0 and *D*=1 or 2. We find that both invariants are nontrivial for the moiré Dirac bands of t-BLG (per spin-valley). One $Z_2$ invariant ((*d-D*) mod 8 = 7 in Table 1) amounts to the quantized Berry phase (0 or π without distinguishing the sign) of the lower band along a loop in the MBZ. The π Berry phase ensures the presence of Dirac points at $K_s$ or $K_s'$ in the MBZ, which is confirmed by our observation of minimum conductivity at the charge neutrality below ~60 K (Figure S1a) and previous measurements by other groups.[32, 40]

| (*d-D*) mod 8 | 0 | 1 | 2 | 3 | 4 | 5 | 6 | 7 |
|---|---|---|---|---|---|---|---|---|
| Invariant | Z | 0 | 0 | 0 | 2Z | 0 | $Z_2$ | $Z_2$ |

Table 1. Topological classification[33] for class AI in the Altland-Zirnbauer table. The cases with the dimensions *d* = 0 and *D* = 1, 2 are relevant for the moiré Dirac bands of t-BLG. Because of the unique $C_{2z}T$ symmetry, the nontrivial $Z_2$ index of the effective dimensions (*d-D*) mod 8 = 7 protects each bulk Dirac point, and the nontrivial $Z_2$ index of the effective dimensions (*d-D*) mod 8 = 6 (Figure 2d) leads to the counter-propagating edge states in each superlattice gap (Figure 2e).

The other $Z_2$ invariant ((*d-D*) mod 8 = 6 in Table 1) over the entire MBZ can be visualized by computing the Wilson loop spectral flow[43] of the two moiré Dirac bands. This spectral flow corresponds to a Wannier-center counterflow (calculated by using circle 1 in Figure 2c) along a loop (circle 2 in Figure 2c) around the MBZ (torus in Figure 2c), and the $Z_2$ invariant characterizes the parity of the counterflow winding. In Figure 2d, the red and cyan traces illustrate the counterflow winding for the two moiré Dirac bands, and the two crossings are symmetry-enforced stable points (Supporting Information Section 5). The nontrivial counterflow, reminiscent of that of 2D TI ,[43] implies the presence of a pair of counter-propagating edge states per spin-valley in the electron superlattice gap (red and cyan curves in Figure 2e). Consistent topological features including similar nontrivial Wilson loop spectral flows have been predicted previously[24-26] based on different approaches. Because of the particle-hole symmetry evidenced in Figure 2a such edge states also exist in the hole superlattice gap. Yet, due to the breaking of $C_{2z}T$ and valley symmetries by t-BLG boundary, the edge states acquire gaps. This is consistent with our observation of non-quantized nonlocal responses at low temperatures; the gapped edge states can form quasi-one-dimensional diffusion channels.

Here we want to emphasize that these gapped topological edge states should be distinguished from those non-topological ones. The non-topological modes, which have an edge origin, inevitably become localized in the presence of substantial edge disorder and roughness, i.e., the case for our devices, and cannot produce any significant non-local transport signals. By sharp contrast, the topological edge states acquire mass terms with different signs in random mesoscopic domains, and the sign reversal at each domain wall yields a 0D zero mode per spin-valley. On statistical average,[44-46] these zero modes weakly couple and form delocalized, diffusive, quasi-1D channels (Figure 2f) that can be responsible for our observed strong, non-quantized, nonlocal signature.

We extend our nonlocal measurements to t-BLG of different twist angles above the magic angle. We observe that the nonlocal responses are universal in all our t-BLG devices with twist angles from ~1.3° to ~1.9°. Figure 3 denotes the nonlocal measurement results from devices D4 to D8. For each twist angle, the red curve plots the measured $R_{NL}$ data recorded at 80 K, whereas the blue curve plots the calculated density of states (DOS) as a function of carrier concentration $n$ (Methods and Supporting Information Section 6). For each twist angle, the pronounced $R_{NL}$ peaks coincide well with the DOS minima, which correspond to the two superlattice gaps separating the moiré

Dirac bands in t-BLG spectra. We note that the moiré Dirac bands of our studied small-angle t-BLG are topologically equivalent since the superlattice gaps do not close as the twist angle decrease from ~2° to the magic angle ~1.1° (Supporting Information Section 6 and Figures S6a-d). Thus, our observation above the magic angle reveals unambiguously the nontrivial band topology of the moiré Dirac bands as a whole near the magic angle. Moreover, the standard etching process (Methods) defined arbitrary edge terminations in our devices. Thus, the existence of strong nonlocal responses in all our t-BLG devices implies the robustness of the observed phenomenon against the edge termination.

We also examine the impact of interlayer displacement field on the nonlocal responses using a dual-gate device (Supporting Information Section 7). The nonlocal responses persist in a wide range of field from $-0.62$ V/nm to $+0.62$ V/nm (Figure S7b). The fact that the nonlocal signal does not vanish at zero field supports our earlier exclusion of inversion symmetry breaking as the origin of the observed nonlocal signals. Moreover, it is also reasonable that a finite field does not remove the nonlocal signal. The field neither breaks the $C_{2z}T$ symmetry nor closes the superlattice gaps (Supporting Information Section 6 and Figures S6i-6l), and thus the presence of delocalized edge modes is intact.

To further elaborate the physical mechanism of the observed nonlocal responses, we measure the length dependence of $R_{NL}$ in a multi-terminal Hall bar device (D9) at 80 K. From the local resistance measurement, we estimate that the twisted angle is ~1.89°. Figure 4a displays three set of $R_{NL}$ data obtained by choosing three channels with different driving and/or probing terminal pairs, as illustrated by the device micrograph in the upper inset in Figure 4a. We notice that the peak positions of different channels are slightly off aligned, which is likely due to the charge inhomogeneity in the sample. Quantitatively, the $R_{NL}$ peak values exponentially decay with the channel length $L$ (lower inset in Figure 4a), and the linear fitting in the semi-log plot yields characteristic diffusion lengths $\lambda_e \approx 1.2$ μm and $\lambda_h \approx 1.4$ μm for the electron- and hole-side, respectively.

The scaling between $R_{NL}$ and $R_L$ is essential to understand the nonlocal transport mechanism. A power-law relation $R_{NL} \sim R_L^\alpha$ is revealed by the Arrhenius plot (Figure 4b) for the data points around $R_{NL}$ ($R_L$) peaks in the range $n \pm \Delta n = -5.54 \pm 0.28$ and $5.54 \pm 0.22$ ($10^{12}\text{cm}^{-2}$) measured in a 1.54° t-BLG device (D10). The fitting parameter $\alpha$ is found to be 3.00 (2.51), 3.37 (2.36), 3.01 (2.17) and 3.50 (2.29) for hole-side (electron-side) responses at 80, 90, 100, 110 K, respectively. This power-law scaling, together with the above revealed diffusion behavior, suggests that $R_{NL} \sim R_L^\alpha e^{-L/\lambda}$. Furthermore, the magnitude of $R_{NL}$ peaks on both electron and hole sides are reduced by ~100 times from 80 to 180 K (Figure 4c). At room temperature, the nonlocal responses can no longer be clearly identified. From the semi-log plot of $R_{NL}$ peak versus $1/T$ (inset in Figure 4c), it is clear that $R_{NL}$ has similar thermal activation behavior as $R_L$ in the superlattice gaps. The activation energies $E_{NL}$ ($E_L$) are extracted to be $157 \pm 20$ meV ($36 \pm 1$ meV) and $129 \pm 16$ meV ($34 \pm 2$ meV) for the electron and hole sides, respectively, with $E_{NL}/E_L \sim 4$. We attribute this deviation from $\alpha$ to the temperature dependence of $\lambda$, since stronger scattering at higher temperature reduces $\lambda$ and effectively enhances $E_{NL}/E_L$.

The observed scaling relation between $R_{NL}$ and $R_L$ in t-BLG is rather unusual. Previously, similar behavior is suggested for the spin Hall effect [47] and seen for the valley Hall effects,[28-31] where the nonlocal transport is captured by a cubic law $R_{NL} \sim R_L^3 \sigma_{xy}^{v,s} e^{-L/\lambda^{v,s}}/\lambda^{v,s}$. Here $\sigma_{xy}^{v,s}$ and $\lambda^{v,s}$ refer to the valley (spin) Hall conductivity and the valley (spin) diffusion length, respectively. In t-BLG, the symmetry-trivialized Berry curvature and negligibly weak spin-orbit coupling imply vanishing $\sigma_{xy}^{v,s}$. However, the gapped yet delocalized edge states forms quasi-one-dimensional diffusion channels and support the nonlocal transport over 1 μm at 80 K. Future measurements through scanning tunneling microscope,[40, 48] scanning superconducting quantum interference,[49, 50] and microwave impedance microscopy[51] may directly image such edge channels in the t-BLG superlattice gaps. Nevertheless, the robustness of observed nonlocal signatures against the interlayer electric field, twist angle, and edge termination not only provides compelling evidence for the nontrivial band topology of t-BLG but also may open a new avenue for creating, engineering, and exploiting moiré quantum information by twistronics and nonlocal means.

## METHODS

### Heterostructure assembly and device fabrication

The hBN encapsulated low-twist-angle bilayer graphene were fabricated using a "tear-and-stack" dry-transfer technique.[5, 32, 35] Graphene and hBN flakes were first exfoliated onto SiO$_2$/Si wafers and examined under optical microscopy and atomic force microscopy. Monolayer graphene and hBN thin flakes of 15 to 30 nm thickness were selected for the heterostructure assembly using a home-made micro-manipulation stage (precision of 0.08°). hBN/t-BLG/hBN heterostructure was assembled on 90 nm-SiO$_2$/Si wafer. All assembly steps were kept below 180 °C to avoid any possible relaxation of the t-BLG into Bernal BLG. We applied standard electron-beam lithography and reactive ion etching (RIE) to create Hall bars of hBN encapsulated t-BLG. A second-round of e-beam lithography and RIE were performed to expose graphene edges for the deposition of contact metals (3 nm Cr/47 nm Au). For the dual-gate device, another layer of hBN (~15 nm) was transferred onto the fabricated device using the aforementioned assembly technique. At last electron-beam lithography and metal deposition (3 nm Cr/37 nm Au) were carried out to form top-gate electrodes.

### Measurements

Electrical characterizations of local and nonlocal responses for device D1, D2, and D3 were performed in a He-3 cryostat. We used standard low-frequency lock-in techniques for the local resistance measurements. For nonlocal responses, we performed direct-current (DC) measurements. A constant current (10nA-100nA) were applied with a Keithley 6221 current source and the nonlocal voltage drop were measured by a Keithley 2182A nanovoltmeter. The DC measurements of local and nonlocal responses for other devices (D4-D11) were carried out in a Lakeshore cryogenic probe station (Model CPX) using an Agilent B1500A semiconductor parameter analyzer. The detailed measurement schemes are discussed in Supporting Information Section 2.

### Theoretical model of t-BLG

We obtain the moiré band structures of t-BLG with small twist angles by using the 2011 Bistritzer-MacDonald model.[4] In this model, the AA and AB tunneling amplitudes are $t_{AA} = t_{AB} = 110$ meV, the Fermi velocity is $v_F = 1 \times 10^6$ m/s, and the flat bands appear at the first magic angle 1.08°.[4] To take into account the corrugation effect due to lattice relaxation, as well as to better match the previous experimental observations, Koshino *et al.* suggested[52] to use $t_{AA}$ = 79.7 meV, $t_{AB}$ = 97.5 meV, and $v_F$ = 7.98 × 10⁵ m/s. In this work, we use these refined parameter values. In our calculations, the momentum cutoff is 6 times of the moiré reciprocal lattice vectors. Namely, the considered area, centered at the *K* point of the original BZ, is 108 times of the area of the first MBZ. To obtain the DOS, we discretize the first MBZ into a mesh with 120,000 points (with the hexagonal symmetry intact) and use 1 meV as the integrated energy interval. The spin-valley degeneracy has been taken into account in DOS calculations. Each moiré band is spin degenerate, since the spin-orbit coupling is negligibly weak in t-BLG. This implies the spin *SU(2)* symmetry. The moiré bands at valley *K'* of the original BZ (not shown) can be obtained by the *T* symmetry from the bands at valley *K*. As evidenced by many graphene experiments, the two valleys are well decoupled in the bulk yet unavoidably coupled near atomic edges. This implies that valley is a good quantum number in the bulk but not near the edges.

**ASSOCIATED CONTENT**

**Supporting Information**

1. Temperature dependence of $R_L$ and nonlocal responses at 1.7 K in 1.68° t-BLG
2. Estimation of the experimental errors in nonlocal measurements
3. Temperature dependence, magneto-transport and low-temperature nonlocal responses for 0.75° t-BLG
4. Temperature dependence, magneto-transport and low-temperature nonlocal responses for 0.42° t-BLG
5. $C_{2z}T$ symmetry, trivial Berry curvature, and nontrivial moiré band topology
6. Twist angle and field dependence of moiré bands
7. Dual-gate mapping of $R_{NL}$ and $R_L$ in device D11
8. Experimental details of all devices presented in this work

Figure S1. Supporting Information transport data for local and nonlocal resistances in device D1 (1.68°).

Figure S2. Estimation of experimental errors in the nonlocal measurements.

Figure S3. Supporting Information transport data for local and nonlocal resistances in device D2 (0.75°).

Figure S4. Transport data for local and nonlocal resistances in device D3 (0.42°).

Figure S5. Moiré Brillouin zone and Wilson loop spectral flow for the two moiré Dirac bands.

Figure S6. Moiré band structures and DOS of t-BLG.

Figure S7. Displacement field dependence of local and nonlocal resistances.

Table S1. Experimental details of all devices presented in this work.

**AUTHOR INFORMATION**

**Corresponding Author**

*Email: fengnian.xia@yale.edu

*Email: zhang@utdallas.edu

*Email: xu.du@stonybrook.edu

**Present Addresses**

[#]Department of Electrical and Electronic Engineering, Southern University of Science and Technology, Shenzhen 518055, China.

**Author Contributions**

C.M. and Q.W. contributed equally to this work. C.M. fabricated the devices. Q.W. performed theoretical calculations. C.M. and S.M. carried out the measurements, with the help from X.C., B.D., S.Y. and C.L. C.M., F.X., F.Z. and X.D. analyzed the data. K.W. and T.T. provided hBN crystals. F.X., F.Z and X.D. co-supervised the project. All authors contributed to the writing of the manuscript.

**Notes**

The authors declare no competing interests.

**ACKNOWLEDGEMENTS**

The work at Yale University is partially supported by the National Science Foundation (NSF) EFRI NewLAW Program and the Office of Naval Research Young Investigator Program (ONR-YIP). The theoretical works (QW and FZ) are supported by Army Research Office (ARO) under Grant Number W911NF-18-1-0416 and partially by the NSF DMREF program under Grant No. DMR-1921581. FZ is grateful to Fengcheng Wu and Adrian Po for valuable discussions. K.W. and T.T. acknowledge support from the Elemental Strategy Initiative conducted by the MEXT, Japan and the CREST (JPMJCR15F3), JST.

**REFERENCES**

(1) Rong, Z. Y.; Kuiper, P. Electronic effects in scanning tunneling microscopy: moire pattern on a graphite surface. *Phys. Rev. B* **1993,** 48, (23), 17427-17431.

(2) Lopes dos Santos, J. M. B.; Peres, N. M. R.; Castro Neto, A. H. Graphene bilayer with a twist: electronic structure. *Phys. Rev. Lett.* **2007,** 99, (25), 256802.

(3) Mele, E. J. Commensuration and interlayer coherence in twisted bilayer graphene. *Phys. Rev. B* **2010,** 81, (16), 161405.

(4) Bistritzer, R.; MacDonald, A. H. Moiré bands in twisted double-layer graphene. *Proc. Natl Acad. Sci. USA* **2011,** 108, (30), 12233-12237.

(5) Kim, K.; Yankowitz, M.; Fallahazad, B.; Kang, S.; Movva, H. C. P.; Huang, S.; Larentis, S.; Corbet, C. M.; Taniguchi, T.; Watanabe, K.; Banerjee, S. K.; LeRoy, B. J.; Tutuc, E. Van der Waals Heterostructures with High Accuracy Rotational Alignment. *Nano Lett.* **2016,** 16, (3), 1989-1995.

(6) Hunt, B.; Sanchez-Yamagishi, J. D.; Young, A. F.; Yankowitz, M.; LeRoy, B. J.; Watanabe, K.; Taniguchi, T.; Moon, P.; Koshino, M.; Jarillo-Herrero, P.; Ashoori, R. C. Massive Dirac Fermions and Hofstadter Butterfly in a van der Waals Heterostructure. *Science* **2013,** 340, (6139), 1427-1430.

(7) Dean, C. R.; Wang, L.; Maher, P.; Forsythe, C.; Ghahari, F.; Gao, Y.; Katoch, J.; Ishigami, M.; Moon, P.; Koshino, M.; Taniguchi, T.; Watanabe, K.; Shepard, K. L.; Hone, J.; Kim, P. Hofstadter's butterfly and the fractal quantum Hall effect in moiré superlattices. *Nature* **2013,** 497, (7451), 598–602.

(8) Ponomarenko, L. A.; Gorbachev, R. V.; Yu, G. L.; Elias, D. C.; Jalil, R.; Patel, A. A.; Mishchenko, A.; Mayorov, A. S.; Woods, C. R.; Wallbank, J. R.; Mucha-Kruczynski, M.; Piot, B. A.; Potemski, M.; Grigorieva, I. V.; Novoselov, K. S.; Guinea, F.; Fal'ko, V. I.; Geim, A. K. Cloning of Dirac fermions in graphene superlattices. *Nature* **2013,** 497, (7451), 594–597.

(9) Seyler, K. L.; Rivera, P.; Yu, H.; Wilson, N. P.; Ray, E. L.; Mandrus, D. G.; Yan, J.; Yao, W.; Xu, X. Signatures of moiré-trapped valley excitons in $MoSe_2/WSe_2$ heterobilayers. *Nature* **2019,** 567, (7746), 66-70.

(10) Jin, C.; Regan, E. C.; Yan, A.; Iqbal Bakti Utama, M.; Wang, D.; Zhao, S.; Qin, Y.; Yang, S.; Zheng, Z.; Shi, S.; Watanabe, K.; Taniguchi, T.; Tongay, S.; Zettl, A.; Wang, F. Observation of moiré excitons in $WSe_2/WS_2$ heterostructure superlattices. *Nature* **2019,** 567, (7746), 76-80.

(11) Tran, K.; Moody, G.; Wu, F.; Lu, X.; Choi, J.; Kim, K.; Rai, A.; Sanchez, D. A.; Quan, J.; Singh, A.; Embley, J.; Zepeda, A.; Campbell, M.; Autry, T.; Taniguchi, T.; Watanabe, K.; Lu, N.; Banerjee, S. K.; Silverman, K. L.; Kim, S.; Tutuc, E.; Yang, L.; MacDonald, A. H.; Li, X. Evidence for moiré excitons in van der Waals heterostructures. *Nature* **2019,** 567, (7746), 71-75.

(12) Alexeev, E. M.; Ruiz-Tijerina, D. A.; Danovich, M.; Hamer, M. J.; Terry, D. J.; Nayak, P. K.; Ahn, S.; Pak, S.; Lee, J.; Sohn, J. I.; Molas, M. R.; Koperski, M.; Watanabe, K.; Taniguchi, T.; Novoselov, K. S.; Gorbachev, R. V.; Shin, H. S.; Fal'ko, V. I.; Tartakovskii, A. I. Resonantly hybridized excitons in moiré superlattices in van der Waals heterostructures. *Nature* **2019,** 567, (7746), 81-86.

(13) Cao, Y.; Fatemi, V.; Demir, A.; Fang, S.; Tomarken, S. L.; Luo, J. Y.; Sanchez-Yamagishi, J. D.; Watanabe, K.; Taniguchi, T.; Kaxiras, E.; Ashoori, R. C.; Jarillo-Herrero, P. Correlated insulator behaviour at half-filling in magic-angle graphene superlattices. *Nature* **2018,** 556, (7699), 80–84.

(14) Cao, Y.; Fatemi, V.; Fang, S.; Watanabe, K.; Taniguchi, T.; Kaxiras, E.; Jarillo-Herrero, P. Unconventional superconductivity in magic-angle graphene superlattices. *Nature* **2018,** 556, (7699), 43–50.

(15) Yankowitz, M.; Chen, S.; Polshyn, H.; Zhang, Y.; Watanabe, K.; Taniguchi, T.; Graf, D.; Young, A. F.; Dean, C. R. Tuning superconductivity in twisted bilayer graphene. *Science* **2019,** 363, (6431), 1059-1064.

(16) Codecido, E.; Wang, Q.; Koester, R.; Che, S.; Tian, H.; Lv, R.; Tran, S.; Watanabe, K.; Taniguchi, T.; Zhang, F.; Bockrath, M.; Lau, C. N. Correlated insulating and superconducting states in twisted bilayer graphene below the magic angle. *Sci. Adv.* **2019,** 5, (9), eaaw9770.

(17) Lu, X.; Stepanov, P.; Yang, W.; Xie, M.; Aamir, M. A.; Das, I.; Urgell, C.; Watanabe, K.; Taniguchi, T.; Zhang, G.; Bachtold, A.; MacDonald, A. H.; Efetov, D. K. Superconductors, orbital magnets and correlated states in magic-angle bilayer graphene. *Nature* **2019,** 574, (7780), 653–657.

(18) Chen, G.; Jiang, L.; Wu, S.; Lyu, B.; Li, H.; Chittari, B. L.; Watanabe, K.; Taniguchi, T.; Shi, Z.; Jung, J.; Zhang, Y.; Wang, F. Evidence of a gate-tunable Mott insulator in a trilayer graphene moiré superlattice. *Nat. Phys.* **2019,** 15, (3), 237-241.

(19) Chen, G.; Sharpe, A. L.; Gallagher, P.; Rosen, I. T.; Fox, E.; Jiang, L.; Lyu, B.; Li, H.; Watanabe, K.; Taniguchi, Jung, T. J.; Shi, Z.; Goldhaber-Gordon, D.; Zhang, Y.; Wang, F. Signatures of tunable superconductivity in a trilayer graphene moiré superlattice. *Nature* **2019,** 572, (7768), 215–219.

(20) Laughlin, R. B. Anomalous quantum Hall effect: an incompressible quantum fluid with fractionally charged excitations. *Phys. Rev. Lett.* **1983,** 50, (18), 1395-1398.

(21) Thouless, D. J.; Kohmoto, M.; Nightingale, M. P.; den Nijs, M. Quantized Hall conductance in a two-dimensional periodic potential. *Phys. Rev. Lett.* **1982,** 49, (6), 405-408.

(22) Zhang, F.; Kane, C. L. Time-reversal-invariant $Z_4$ fractional Josephson effect. *Phys. Rev. Lett.* **2014,** 113, (3), 036401.

(23) Kane, C. L.; Mele, E. J. $Z_2$ Topological order and the quantum spin Hall effect. *Phys. Rev. Lett.* **2005,** 95, (14), 146802.

(24) Song, Z.; Wang, Z.; Shi, W.; Li, G.; Fang, C.; Bernevig, B. A. All magic angles in twisted bilayer graphene are topological. *Phys. Rev. Lett.* **2019,** 123, (3), 036401.

(25) Po, H. C.; Zou, L.; Senthil, T.; Vishwanath, A. Faithful tight-binding models and fragile topology of magic-angle bilayer graphene. *Phys. Rev. B* **2019,** 99, (19), 195455.

(26) Ahn, J.; Park, S.; Yang, B.-J. Failure of Nielsen-Ninomiya theorem and fragile topology in two-dimensional systems with space-time inversion symmetry: application to twisted bilayer graphene at magic angle. *Phys. Rev. X* **2019,** 9, (2), 021013.

(27) Roth, A.; Brüne, C.; Buhmann, H.; Molenkamp, L. W.; Maciejko, J.; Qi, X.-L.; Zhang, S.-C. Nonlocal transport in the quantum spin Hall state. *Science* **2009,** 325, (5938), 294-297.

(28) Gorbachev, R. V.; Song, J. C. W.; Yu, G. L.; Kretinin, A. V.; Withers, F.; Cao, Y.; Mishchenko, A.; Grigorieva, I. V.; Novoselov, K. S.; Levitov, L. S.; Geim, A. K. Detecting topological currents in graphene superlattices. *Science* **2014,** 346, (6208), 448-451.

(29) Sui, M.; Chen, G.; Ma, L.; Shan, W.-Y.; Tian, D.; Watanabe, K.; Taniguchi, T.; Jin, X.; Yao, W.; Xiao, D.; Zhang, Y. Gate-tunable topological valley transport in bilayer graphene. *Nat. Phys.* **2015,** 11, (12), 1027–1031.

(30) Shimazaki, Y.; Yamamoto, M.; Borzenets, I. V.; Watanabe, K.; Taniguchi, T.; Tarucha, S. Generation and detection of pure valley current by electrically induced Berry curvature in bilayer graphene. *Nat. Phys.* **2015,** 11, (12), 1032–1036.

(31) Wu, Z.; Zhou, B. T.; Cai, X.; Cheung, P.; Liu, G.-B.; Huang, M.; Lin, J.; Han, T.; An, L.; Wang, Y.; Xu, S.; Long, G.; Cheng, C.; Law, K. T.; Zhang, F.; Wang, N. Intrinsic valley Hall transport in atomically thin $MoS_2$. *Nat. Commun.* **2019,** 10, (1), 611.

(32) Cao, Y.; Luo, J. Y.; Fatemi, V.; Fang, S.; Sanchez-Yamagishi, J. D.; Watanabe, K.; Taniguchi, T.; Kaxiras, E.; Jarillo-Herrero, P. Superlattice-induced insulating states and valley-protected orbits in twisted bilayer graphene. *Phys. Rev. Lett.* **2016,** 117, (11), 116804.

(33) Teo, J. C. Y.; Kane, C. L. Topological defects and gapless modes in insulators and superconductors. *Phys. Rev. B* **2010,** 82, (11), 115120.

(34) Sharpe, A. L.; Fox, E. J.; Barnard, A. W.; Finney, J.; Watanabe, K.; Taniguchi, T.; Kastner, M. A.; Goldhaber-Gordon, D. Emergent ferromagnetism near three-quarters filling in twisted bilayer graphene. *Science* **2019,** 365, (6453), 605-608.

(35) Wang, L.; Meric, I.; Huang, P. Y.; Gao, Q.; Gao, Y.; Tran, H.; Taniguchi, T.; Watanabe, K.; Campos, L. M.; Muller, D. A.; Guo, J.; Kim, P.; Hone, J.; Shepard, K. L.; Dean, C. R. One-dimensional electrical contact to a two-dimensional material. *Science* **2013,** 342, (6158), 614-617.

(36) Brüne, C.; Roth, A.; Novik, E. G.; König, M.; Buhmann, H.; Hankiewicz, E. M.; Hanke, W.; Sinova, J.; Molenkamp, L. W. Evidence for the ballistic intrinsic spin Hall effect in HgTe nanostructures. *Nat. Phys.* **2010,** 6, (6), 448–454.

(37) Abanin, D. A.; Morozov, S. V.; Ponomarenko, L. A.; Gorbachev, R. V.; Mayorov, A. S.; Katsnelson, M. I.; Watanabe, K.; Taniguchi, T.; Novoselov, K. S.; Levitov, L. S.; Geim, A. K. Giant nonlocality near the Dirac point in graphene. *Science* **2011,** 332, (6027), 328-330.

(38) Balakrishnan, J.; Kok Wai Koon, G.; Jaiswal, M.; Castro Neto, A. H.; Özyilmaz, B. Colossal enhancement of spin–orbit coupling in weakly hydrogenated graphene. *Nat. Phys.* **2013,** 9, (5), 284-287.

(39) Mishchenko, A.; Cao, Y.; Yu, G. L.; Woods, C. R.; Gorbachev, R. V.; Novoselov, K. S.; Geim, A. K.; Levitov, L. S. Nonlocal response and anamorphosis: the case of few-layer black phosphorus. *Nano Lett.* **2015,** 15, (10), 6991-6995.

(40) Kim, K.; DaSilva, A.; Huang, S.; Fallahazad, B.; Larentis, S.; Taniguchi, T.; Watanabe, K.; LeRoy, B. J.; MacDonald, A. H.; Tutuc, E. Tunable moiré bands and strong correlations in small-twist-angle bilayer graphene. *Proc. Natl Acad. Sci. USA* **2017,** 114, (13), 3364-3369.

(41) Polshyn, H.; Yankowitz, M.; Chen, S.; Zhang, Y.; Watanabe, K.; Taniguchi, T.; Dean, C. R.; Young, A. F. Large linear-in-temperature resistivity in twisted bilayer graphene. *Nat. Phys.* **2019,** 15, (10), 1011-1016.

(42) Zhang, F. Brought to light. *Nat. Phys.* **2017,** 14, (2), 111–113.

(43) Yu, R.; Qi, X. L.; Bernevig, A.; Fang, Z.; Dai, X. Equivalent expression of $Z_2$ topological invariant for band insulators using the non-Abelian Berry connection. *Phys. Rev. B* **2011,** 84, (7), 075119.

(44) Mong, R. S.; Bardarson, J. H.; Moore, J. E. Quantum transport and two-parameter scaling at the surface of a weak topological insulator. *Phys. Rev. Lett.* **2012,** 108, (7), 076804.

(45) Fu, L.; Kane, C. L. Topology, delocalization via average symmetry and the symplectic anderson transition. *Phys. Rev. Lett.* **2012,** 109, (24), 246605.

(46) Ringel, Z.; Kraus, Y. E.; Stern, A. Strong side of weak topological insulators. *Phys. Rev. B* **2012,** 86, (4), 045102.

(47) Abanin, D. A.; Shytov, A. V.; Levitov, L. S.; Halperin, B. I. Nonlocal charge transport mediated by spin diffusion in the spin Hall effect regime. *Phys. Rev. B* **2009,** 79, (3), 035304.

(48) Huang, S.; Kim, K.; Efimkin, D. K.; Lovorn, T.; Taniguchi, T.; Watanabe, K.; MacDonald, A. H.; Tutuc, E.; LeRoy, B. J. Topologically Protected Helical States in Minimally Twisted Bilayer Graphene. *Phys. Rev. Lett.* **2018,** 121, (3), 037702.

(49) Nowack, K. C.; Spanton, E. M.; Baenninger, M.; König, M.; Kirtley, J. R.; Kalisky, B.; Ames, C.; Leubner, P.; Brüne, C.; Buhmann, H.; Molenkamp, L. W.; Goldhaber-Gordon, D.; Moler, K. A. Imaging currents in HgTe quantum wells in the quantum spin Hall regime. *Nat. Mater.* **2013,** 12, (9), 787–791.

(50) Spanton, E. M.; Nowack, K. C.; Du, L.; Sullivan, G.; Du, R.-R.; Moler, K. A. Images of edge current in InAs/GaSb quantum wells. *Phys. Rev. Lett.* **2014,** 113, (2), 026804.

(51) Shi, Y.; Kahn, J.; Niu, B.; Fei, Z.; Sun, B.; Cai, X.; Francisco, B. A.; Wu, D.; Shen, Z.-X.; Xu, X.; Cobden, D. H.; Cui, Y.-T. Imaging quantum spin Hall edges in monolayer $WTe_2$. *Sci. Adv.* **2019,** 5, (2), eaat8799.

(52) Koshino, M.; Yuan, N. F. Q.; Koretsune, T.; Ochi, M.; Kuroki, K.; Fu, L. Maximally localized wannier orbitals and the extended Hubbard model for twisted bilayer graphene. *Phys. Rev. X* **2018,** 8, (3), 031087.

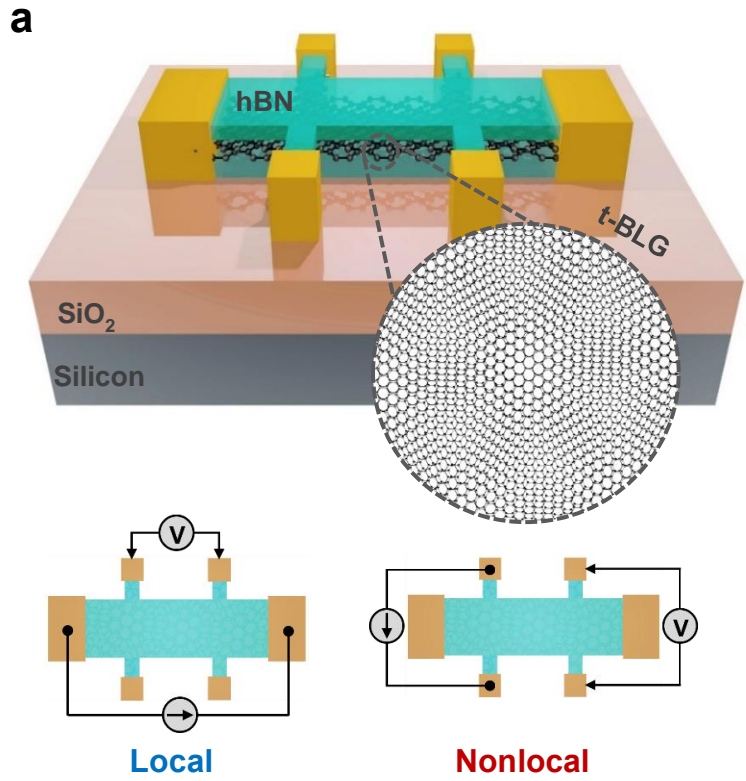

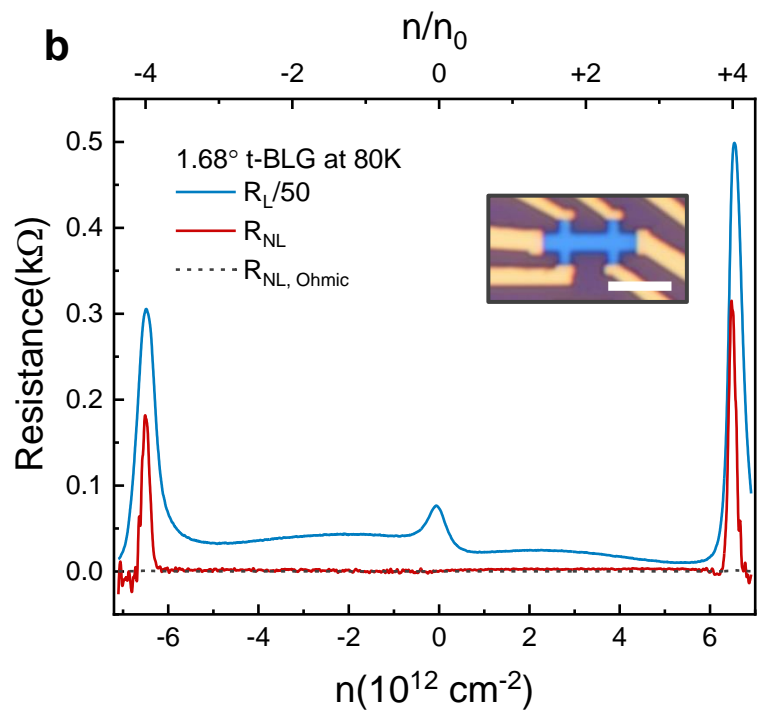

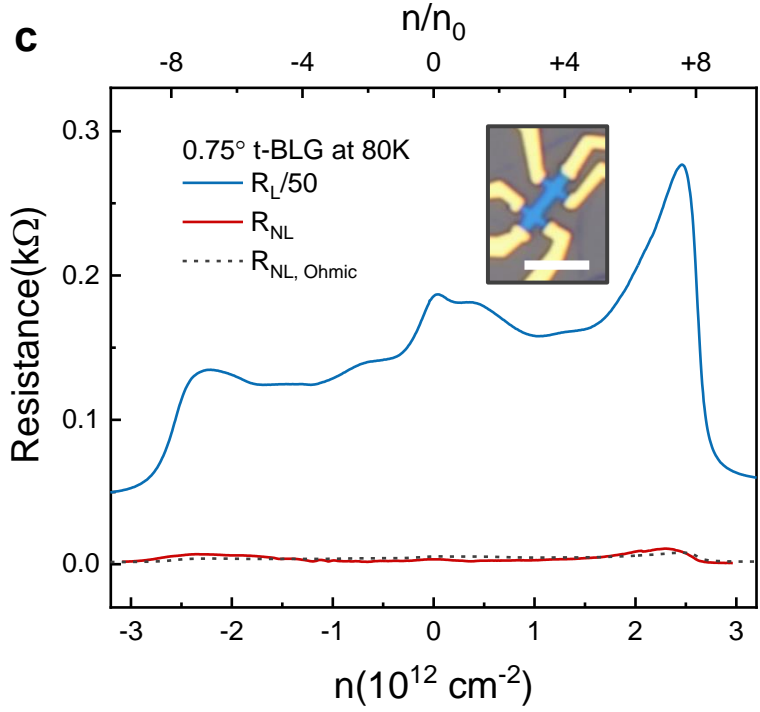

**Figure 1.** Local and nonlocal transport properties in twist bilayer graphene. (a) Top: schematic of a typical single-gate device with t-BLG channel. The carrier concentration $n$ is tuned by a global silicon back gate. Inset: Schematic moiré pattern of t-BLG consisting two layers of monolayer graphene with a twist. Bottom: Schematics of local and nonlocal measurement schemes. (b) Local and nonlocal resistances measured in a t-BLG device (D1) with a 1.68° twist angle at 80 K. The three local resistance maxima correspond to the Dirac points $n/n_0 = 0$ (at charge neutrality) and the superlattice gaps $n/n_0 = \pm 4$ (at $n \approx \pm 6.50 \times 10^{12}$ cm$^{-2}$). Pronounced nonlocal responses are observed inside the superlattice gaps. Inset: Optical image of device D1. Scale bar: 4 μm. (c) Local and nonlocal resistances measured in another t-BLG device (D2) with a 0.75° twist angle at 80 K. Filling factor $n/n_0$ is determined by the low temperature magneto-transport (Figure S3b). No significant nonlocal responses are observed at $n/n_0 = \pm 4$ ($n \approx \pm 1.31 \times 10^{12}$ cm$^{-2}$) due to the closure of the superlattice gaps. Inset: Optical image of device D2. Scale bar: 4 μm.

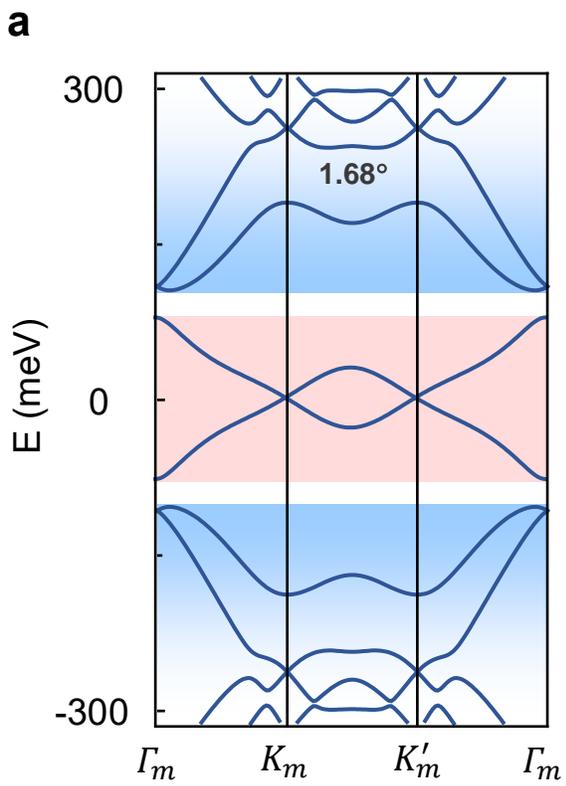
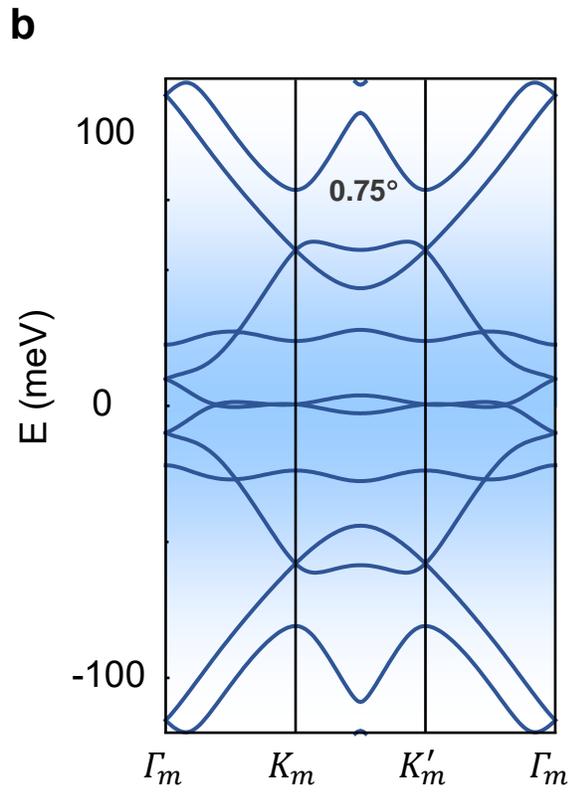
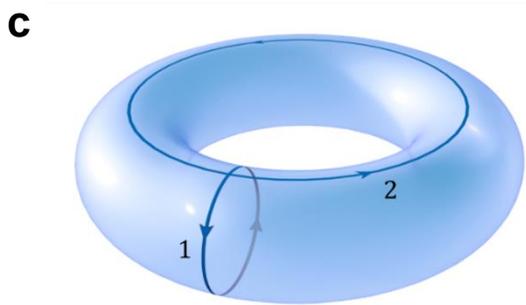
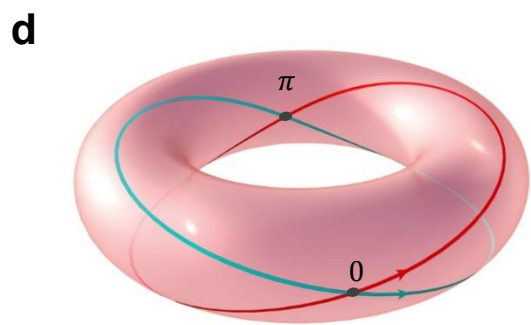
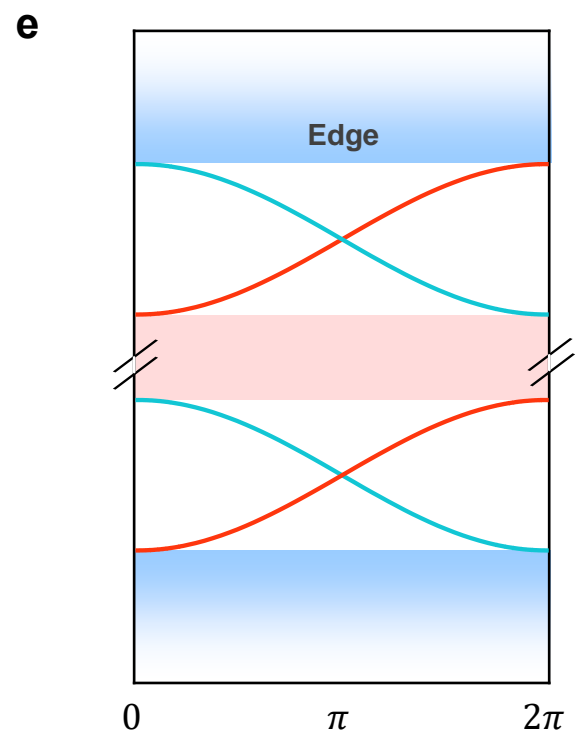

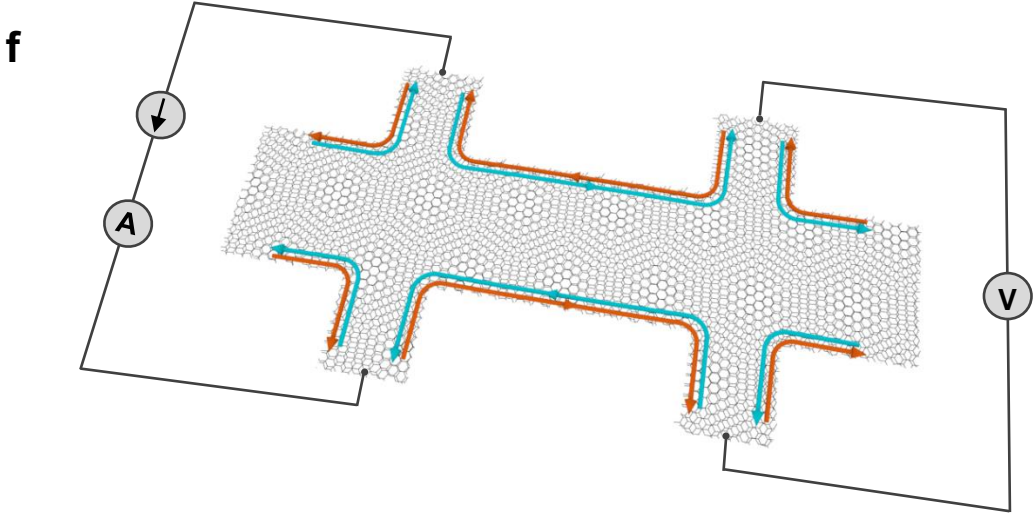

**Figure 2.** Band structure, band topology, and delocalized edge modes of twisted bilayer graphene. (a) The calculated moiré band structure of 1.68° t-BLG. Superlattice gaps are denoted by the white regions while two lowest moiré Dirac bands with a nontrivial $Z_2$ index are denoted by the red shaded region. (b) The calculated moiré band structure of 0.75° t-BLG. Superlattice gaps are absent in this t-BLG system. (c) Schematic torus illustrating the first moiré Brillouin zone. Two Wannier centers of the two moiré Dirac bands along circle 1 are computed, and their counterflow along circle 2 is depicted in (d). (d) Red and cyan traces illustrating the nontrivial Wannier-center counterflow winding, and the crossings labeled by 0 and $\pi$ are symmetry-enforced stable points. (e) Illustration of the counter-propagating edge states per spin-valley (red and cyan curves) in the superlattice bandgaps. The edge states acquire gaps due to the broken $C_{2z}T$ and valley symmetries at t-BLG device boundaries. (f) Schematic of the nonlocal detection of the two counter-propagating edge states in a t-BLG Hall bar.

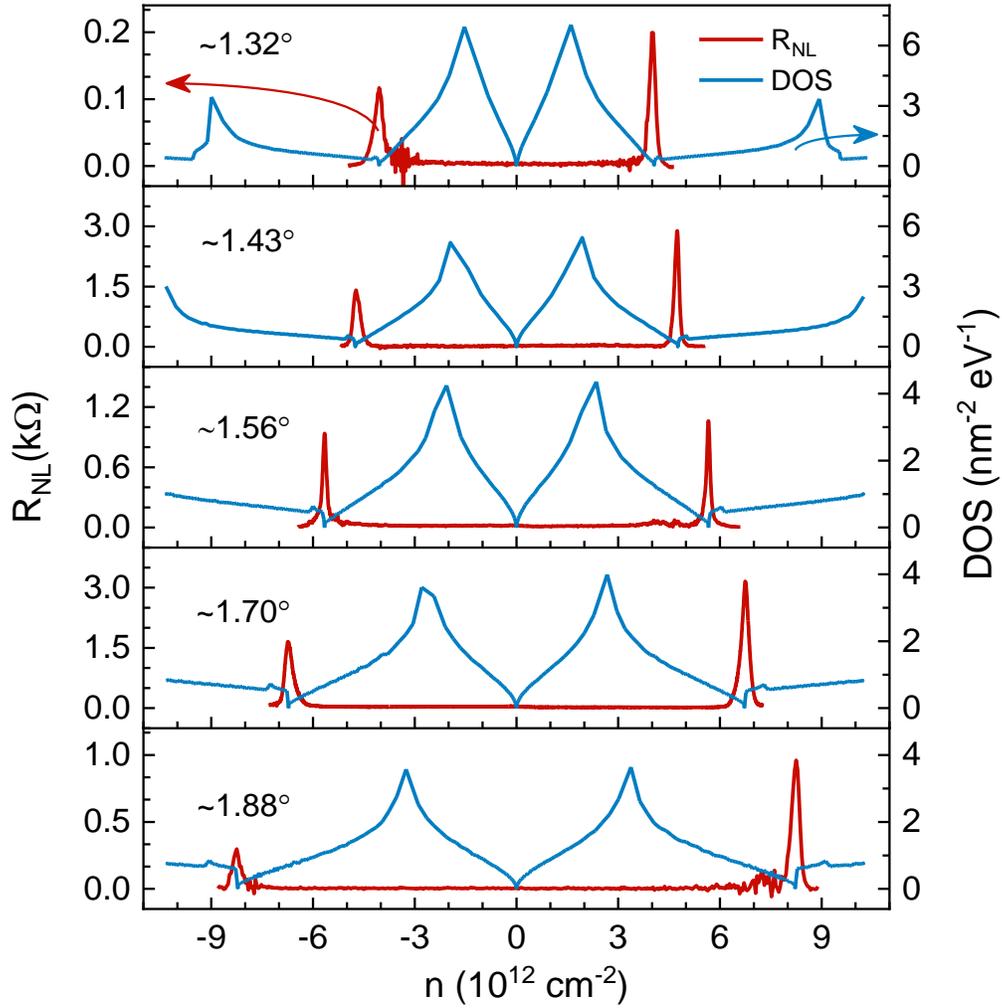

**Figure 3.** Twist angle dependence of nonlocal resistance and DOS. Measured nonlocal resistance (red curves) and calculated density of states (DOS) as functions of $n$ measured in various t-BLG devices (D4 to D8) with different twist angles from ~1.3° to ~1.9° at 80 K (See Supporting Information Section 8 for device details). Nonlocal resistance peaks appear at the DOS minima corresponding to the superlattice bandgaps in each t-BLG.

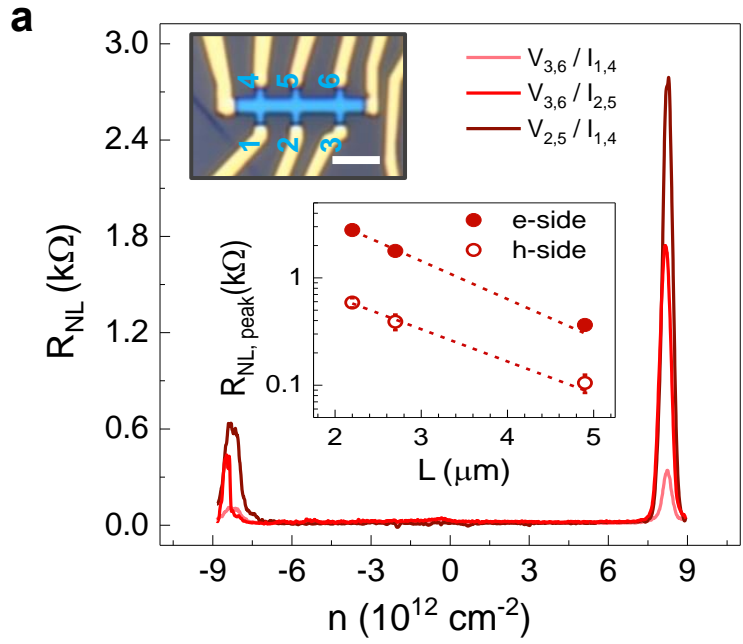

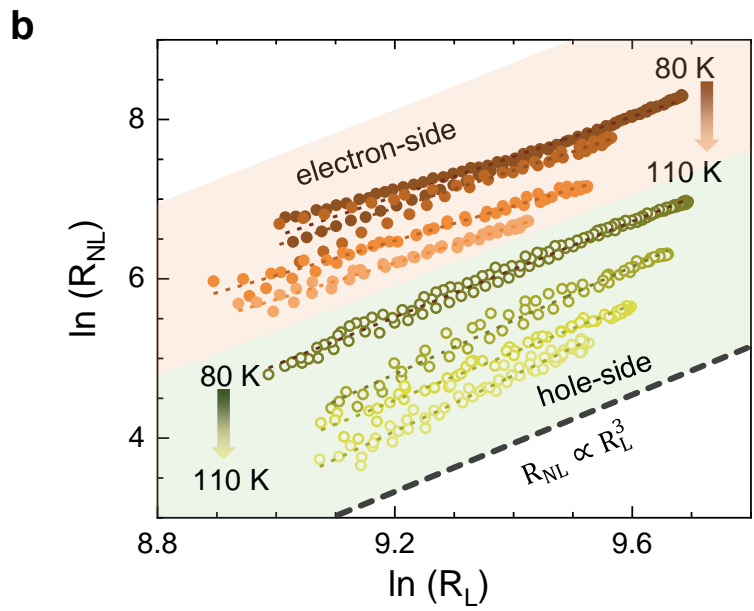

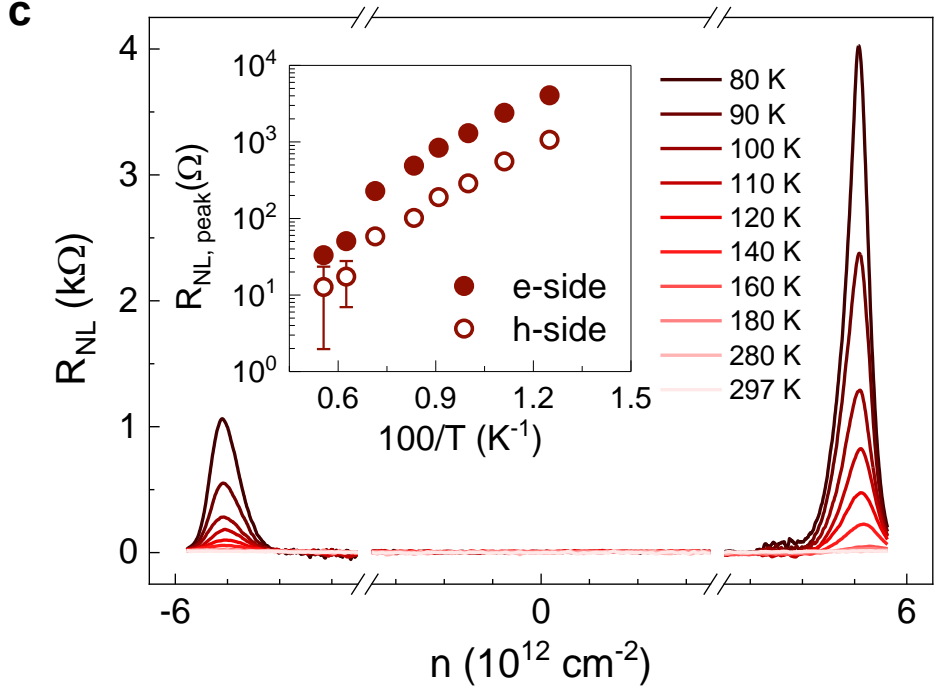

**Figure 4.** Length and temperature dependence of nonlocal resistance. (a) Length dependence of nonlocal responses in a multiterminal device (D9) measured at 80 K. Three sets of $R_{NL}$ are measured by driving current $I$ and probing voltage drop $V$ using different pairs of terminals as indicated in the figure legend. Top left inset: Optical image of D9. Scale bar: 3 μm. Middle inset: Semi-log plot of $R_{NL}$ peaks versus channel length $L$. The nonlocal responses exponentially decay with $L$. Solid points and circles represent the electron- and hole-side $R_{NL}$ peak values, respectively. The fittings (dashed lines) yield diffusion lengths of 1.2 and 1.4 μm for electron and hole sides, respectively. (b) Scaling relation between $R_{NL}$ and $R_L$ measured in D10 at different temperatures. Data points around the $R_{NL}(R_L)$ peaks are extracted in the range $n \pm \Delta n = 5.54 \pm 0.22$ and $-5.54 \pm 0.28$ ($10^{12}$cm$^{-2}$) for the electron (orange shaded region) and hole (green shaded region) sides. A linear relation between $\ln R_{NL}$ and $\ln R_L$ gives a slope 3.00 (2.51), 3.37 (2.36), 3.01 (2.17) and 3.50 (2.29) for hole-side (electron-side) responses at 80, 90, 100, and 110 K, respectively. (c) Temperature dependence of $R_{NL}$ from 80 K up to room temperature. Inset: Semi-log plot of $R_{NL}$ peak values versus $100/T$. Above 80 K, $R_{NL}$ also exhibits thermal activation behavior with the activation energies extracted to be $157 \pm 20$ meV and $129 \pm 16$ meV for the electron and hole sides, respectively.